\documentclass[twocolumn,amsmath,amssymb,prb]{revtex4}

\usepackage[pdftex]{graphicx} 
\usepackage{epstopdf}
\usepackage{verbatim}
\usepackage{color}

\newcommand{\bra}[1]{\ensuremath{\langle#1|}}
\newcommand{\ket}[1]{\ensuremath{|#1\rangle}}
\newcommand{\bracket}[2]{\ensuremath{\langle#1|#2\rangle}}

\newcommand{\approxlt}{ \,{\scriptstyle\lesssim} \,}

\newcommand{\be}{\begin{equation}}
\newcommand{\ee}{\end{equation}}

\begin{document}

\title{Studying Two Dimensional Systems With the Density Matrix Renormalization Group}

\author{E.M. Stoudenmire} 
\affiliation{Department of Physics and Astronomy, University of California, Irvine, CA 92697}
\author{Steven R. White}
\affiliation{Department of Physics and Astronomy, University of California, Irvine, CA 92697}

\date{\today}

\begin{abstract}
Despite a computational effort that scales exponentially with the system width, the
Density Matrix Renormalization Group (DMRG) method is
one of the most powerful numerical methods for studying two dimensional quantum lattice
systems. Reviewing past applications of DMRG in 2D demonstrates its
success in treating a wide variety of problems, although it remains underutilized in this setting. 
We present techniques for performing cutting edge 2D DMRG studies including methods for 
ensuring convergence, extrapolating finite-size data and extracting gaps and excited states.
Finally, we compare the current performance of a recently developed tensor network method to 2D DMRG.
\end{abstract}

\maketitle

\section{Introduction}

The behavior of quantum many-body lattice systems depends strongly on their dimensionality.  
Mean field theory---where the behavior at each site 
is determined self-consistently by the average influence of all its neighbors---works best when there are 
many neighbors, that is, in higher dimensions.  
Three dimensional systems can often be well understood using
mean-field or semi-classical approaches, with quantum fluctuations acting as a 
minor correction.\cite{Bergman:2007,Castelnovo:2008} 
In one dimension, mean field theory usually fails, 
and exotic behavior driven by strong quantum fluctuations, such as spin-charge
separation, is the norm. 
By using powerful analytical and numerical approaches
developed over the last few decades, it is possible to determine the properties 
of 1D systems to high accuracy.

Two dimensional systems often have substantial quantum fluctuations, causing mean-field 
approaches to fail. Yet geometrical constraints are
much more relaxed than in 1D, allowing many more phases to exist. In some
ways, geometry specifically selects 2D as being the most interesting: for
example, in 2D a pair of particles can circle one another, unlike in 1D; but different
numbers of revolutions (and their signs) are topologically distinct, unlike in three
or more dimensions.  This allows for the possibility---in 2D alone---of elementary
excitations of a system's ground state which are anyons, particles that are neither
bosons nor fermions. Currently, two dimensional systems are at the heart of quantum condensed
matter physics, with a large fraction of researchers focusing their efforts on
key families of two dimensional systems, including the high temperature superconducting
cuprates,\cite{Dagotto:1994,Kastner:1998} quantum Hall systems\cite{Stormer:1999} and frustrated magnets, 
especially those that could host spin liquids.\cite{Balents:2010,Yamashita:2009,Yamashita:2010,Yan:2011} 

\begin{figure}[tp]
\includegraphics[width=\columnwidth]{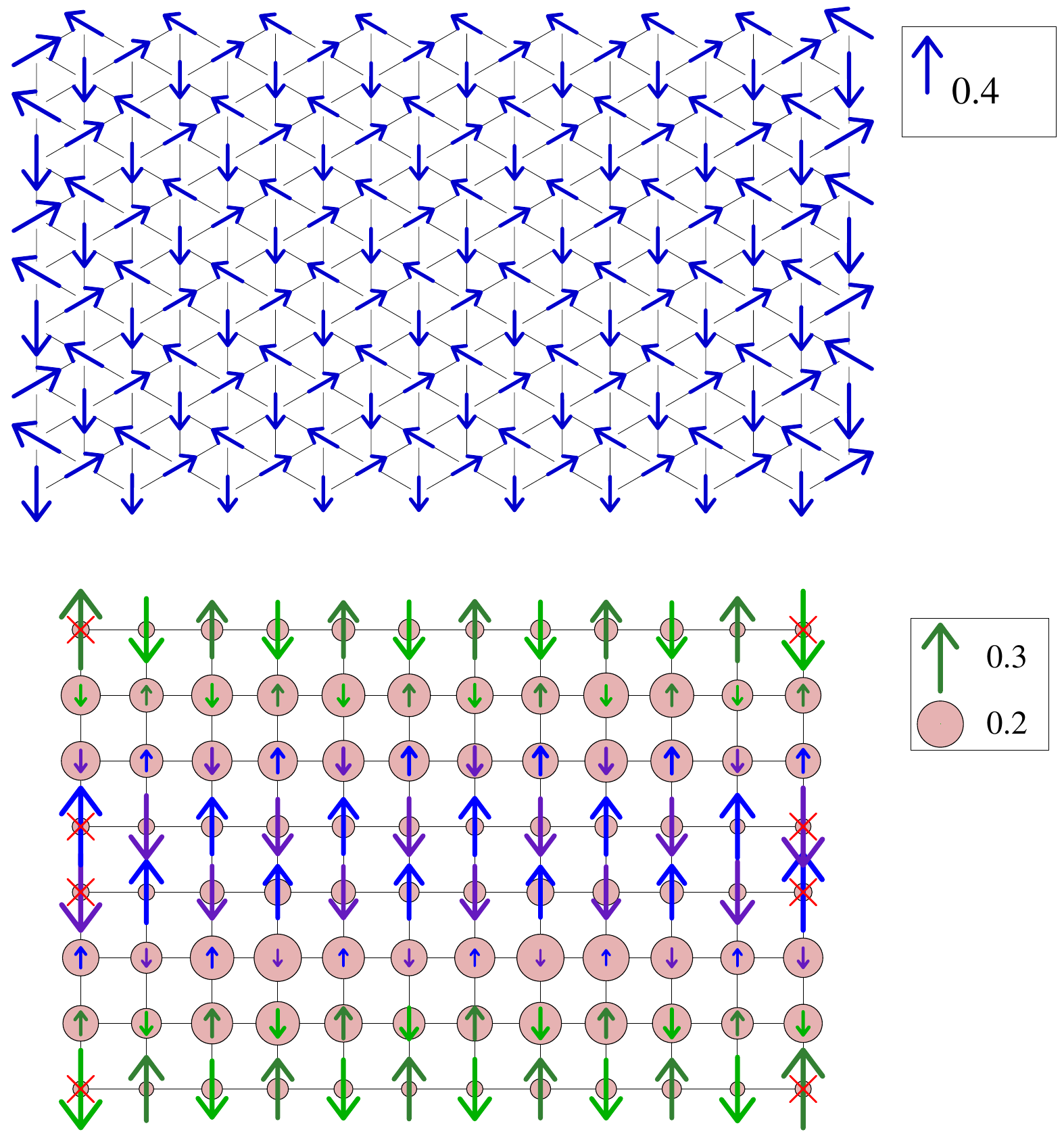}
\caption{
Results of DMRG simulations on 2D systems with either frustration or mobile fermions.
Upper panel: local values of $\langle \vec S\rangle$ for a triangular Heisenberg model.
The 120$^\circ$ antiferromagnetic order is ``pinned'' by applied magnetic fields on
left and right edges; the top and bottom edges are connected to make a cylindrical geometry.
Lower panel:  spin and hole densities
for a $t$-$J$ cluster which exhibits both stripes and pairing. Cylindrical BCs are used, and pinning with both
magnetic and proximity effect pair fields was used on the left and right open ends.}
\label{fig:tri_tJ}
\end{figure}

Most of the analytical and numerical techniques that often work in one or three dimensions
fail in 2D. For example, analytic techniques based on the Bethe ansatz 
(an exact wavefunction for certain 1D systems) or on conformal field theory are specific to 1D.\cite{Giamarchi:2003}
Another example is quantum Monte Carlo, one of the most powerful numerical methods.\cite{Sandvik:2007,Isakov:2011}  
Classical Monte Carlo methods work in any dimension, but the path-integral formulation needed for
quantum Monte Carlo introduces ``probabilities" that can be negative; this is known as
the sign problem.\cite{White:1989,Troyer:2005}
Quantum Monte Carlo is of limited usefulness for systems with
a sign problem. Although the presence of the sign problem does not correspond strictly
to dimensionality, 1D systems with nearest neighbor interactions do not have a
a sign problem, while 2D or 3D systems with either mobile fermions or frustrating
spin interactions do have a sign problem.  ``Solving the sign problem'' is sometimes
called one of the great challenges of condensed matter physics. Narrowly, solving
the sign problem means fixing quantum Monte Carlo methods in some way, but we can
also take a broader view: can we find any numerical method that can solve broad
classes of frustrated or fermionic quantum systems in 2D or 3D?

The density matrix renormalization group (DMRG) is a numerical approach designed
for one dimension and which has become the most powerful known numerical method in 1D.\cite{White:1992,White:1993a}
The subject of this article is the application of DMRG to 2D.  It is more difficult to use
DMRG in 2D, and the results are much less accurate than in 1D.  Nevertheless, the
lack of alternative approaches (when there is a sign problem) makes DMRG one of
the most powerful current methods for certain 2D systems.  It is a key goal of this article to present the tricks
and techniques that are essential for the efficient treatment of 2D systems.

Unfortunately, DMRG has been underutilized in this context. One reason may be the
idea that since DMRG has a computational effort scaling exponentially with the width of
the system, it is not useful. However, this same argument would imply that exact diagonalization 
(solving for the ground state eigenvector utilizing the complete Hilbert space of a finite cluster)
would not be useful even in 1D! In fact, exact diagonalization is often useful even in 2D.\cite{Luscher:2009,Poilblanc:2010} 
The exponential scaling of DMRG in 2D depends only upon the width, not the total
number of sites like in exact diagonalization, and one often finds a fairly modest
coefficient governing the exponential.  One finds that even in
sign-problem-free models, DMRG can make predictions of similar quality to Quantum Monte Carlo.\cite{White:2007}

Figure 1 shows two examples of results from recent 2D DMRG calculations.
The system widths shown are close to the state of the art in terms of maximum
system sizes reachable; but using the techniques discussed below in Section \ref{sec:techniques},
it is often possible to extrapolate such DMRG results to the infinite 2D limit quite reliably.
DMRG makes up for its size limitations even further by providing access to the entire many-body wavefunction,
making it possible to compute essentially any ground state observable.

To understand how DMRG works, observe that while exact diagonalization describes
the wavefunction in a complete, exponentially large basis, some of these coefficients
and basis functions are much more significant than others.  The idea of 
making a systematic approximation by truncating
this basis is quite old, but in its simplest form
it is not very effective for strongly correlated systems.  In DMRG, one first rotates the
basis so as to make the truncation much more accurate---in the rotated basis,
only a few of the states are needed to represent the ground state, while the rest can
be discarded.  A rotation involving the whole Hilbert space would be very inefficient;
instead, DMRG uses many rotations focused on a few sites at a time, generating global
rotations via  a sweeping procedure through all the sites of the lattice.
The result is a wavefunction written in a particular form, called a ``matrix product state"
or MPS. An alternative viewpoint for DMRG is to start with the MPS form as a variational
ansatz, then optimize all its coefficients. From this viewpoint, DMRG is an
extremely efficient method for optimizing the coefficients.\cite{White:1992,Schollwoeck:2005}
The basis used by DMRG is the optimal basis within a certain framework;\cite{Hastings:2007} in
recent years it has been understood to be a straightforward consequence of
the Schmidt decomposition of quantum information.

Because DMRG is so efficient, it has dominated numerical research into strongly 
correlated 1D systems. One of the first applications of DMRG was calculating
the excitation gap in the $S=1$ Heisenberg chain to very high accuracy at a time when its
existence remained controversial.\cite{White:1993}
Since then, not only has the original DMRG algorithm improved but its flexibility has
allowed many extensions. For example, DMRG is very useful for studying time-dependent
Hamiltonians\cite{Schollwoeck:2006} and finite-temperature systems.\cite{White:2009,Stoudenmire:2010}
Following the realization that DMRG is actually based upon matrix product states,\cite{Ostlund:1995}
the method has become highly influential within the quantum information community, leading
to the development of a new breed of algorithms based on tensor network states.\cite{Jiang:2008t,Evenbly:2010,Wang:2011}
These developments have in turn accelerated the power and flexibility of the original DMRG method. For example,
it is now possible to use DMRG to reliably determine the ground state of \emph{infinite} 1D systems.\cite{McCulloch:2008}

While the hope is that tensor network state approaches
will eventually overtake DMRG in two dimensional
studies (see Section \ref{sec:tns}), in the meantime it remains one of the most powerful and well
controlled methods for simulating models with a sign problem. DMRG can even be
useful for simulating sign-problem free models since it provides full
access to the many-body wavefunction.\cite{Kallin:2011}

In the following section we review past applications of DMRG to 2D systems. 
In Section \ref{sec:techniques}---a discussion of techniques for 2D---we assume a working knowledge of
DMRG for 1D systems; for a detailed introduction see, for example, Schollw\"ock,\cite{Schollwoeck:2005,Schollwoeck:2011}
Hallberg,\cite{Hallberg:2006} Noack\cite{Noack:2005} and White.\cite{White:1993a}

\section{Applications of Two Dimensional DMRG}

DMRG has been successfully used to study a wide variety of 2D systems - a
testament to the flexibility of the method. We briefly survey some prior
studies below, focusing especially on those which break new ground by
introducing techniques or characterizing especially interesting systems. The
hope is to convey an understanding of what is possible with current methods and
to motivate the investigation of systems that are good candidates for a DMRG
study. 

One of the first applications of DMRG was in the area of magnetism.\cite{White:1993} 
Magnetic models, such as the Heisenberg model, are ideally suited for DMRG because they can
exhibit interesting phases with only modest entanglement and rarely suffer from
large differences in energy scales.  Among the first 2D magnetic models to be
studied with DMRG was the frustrated $S=1/2$ Heisenberg antiferromagnet on the CAVO
lattice.\cite{White:1996}  This study helped establish the usefulness of DMRG
for 2D systems and introduced the wavefunction acceleration technique,
improving performance by up to two orders of magnitude.  Since then, 2D DMRG
has been used to study the frustrated nearest-neighbor Heisenberg model on the
triangular \cite{Yoshikawa:2004,White:2007,Weng:2006} and kagome lattices
\cite{Jiang:2008,Yan:2011} as well as models frustrated by further neighbor or
multi-spin interactions.\cite{Capriotti:2004b,Jiang:2009b,Sheng:2009,Block:2010,Jiang:2011} 
Two dimensional DMRG is powerful enough to be a good option for studying
unfrustrated spin models too.\cite{White:1994,Croo-de-Jongh:1998,Zhao:2006} One recent
study obtained an estimate for the magnetization of the square lattice
Heisenberg model competitive with the best published Quantum Monte Carlo
results.\cite{White:2007}

An area where 2D DMRG has been especially fruitful is in the search for quantum spin liquids---magnetic systems that break no 
symmetries down to $T=0$. Many realistic, one-dimensional spin models have disordered ground states, but the search for realistic 2D spin liquids
remains a challenge.\cite{Balents:2010} To date, DMRG has been used to identify two-dimensional spin liquid phases stabilized by 
anisotropic,\cite{White:1994,Weng:2006} further neighbor \cite{Capriotti:2004b,Jiang:2009b} and multi-spin \cite{Sheng:2009,Block:2010} interactions, 
yet definitive evidence of a short-range, isotropic spin model whose ground state breaks no symmetries has been lacking.
However, DMRG simulations of the Heisenberg antiferromagnet on the kagome lattice now show strong evidence for a spin liquid 
ground state.\cite{Yan:2011} 
2D DMRG has also been used recently to study perturbations to the `unrealistic' Kitaev honeycomb model that
may actually describe a certain limit of the layered magnet Na$_2$IrO$_3$.\cite{Jiang:2011}

Another set of systems studied extensively with 2D DMRG are the doped $t-J$ and
Hubbard models.  DMRG is an attractive option for simulating models of this
type because it can deal with arbitrary doping in an unbiased way.  A common
thread running through the DMRG literature on the $t-J$ and Hubbard models is
the formation of stripes - charge density waves separating regions of
phase-shifted antiferromagnetic order.  After stripes were found to occur
spontaneously in four-leg $t-J$ ladders, \cite{White:1997} they were also found
to occur in systems up to width eight.\cite{White:1998c} Follow-up studies
have examined properties of stripes more closely, such as their doping and
interaction energy \cite{White:1998} and the effects of further-neighbor
interactions \cite{Tohyama:1999,White:1999} and anisotropies \cite{Kampf:2001}
likely present in real materials.  It was suggested early on
that the stripes observed experimentally are due to frustrated phase separation, \cite{Emery:1990} but evidence from DMRG
favors the idea that stripes are caused by local competition between hopping and exchange,
without the need for long range Coulomb interactions.\cite{White:2000}  
Two-dimensional DMRG results may also shed light on the
pairing mechanism of the cuprates.\cite{Scalapino:2001}

One study which especially highlighted the reliability of DMRG results for 2D $t-J$ models
was Chernyshev et al.'s comparison of 2D DMRG results to predictions from a self-consistent Green function theory 
of holes in an antiferromagnetic background. Working with a modified $t-J$ model where $J$ is replaced by
an Ising $J_z$ interaction, the two methods show remarkable correspondence in their predictions of energies and hole distributions, 
especially for the case of a single hole.\cite{Chernyshev:2002,Chernyshev:2002b}
Further DMRG investigations of the $t-J$ model have examined the existence of checkerboard order, \cite{White:2004} edge states of holes
in nanosystems \cite{Chernyshev:2005} and the competition between stripes and pairing.\cite{White:2009b}
DMRG has also been used to demonstrate a striped phase in the Hubbard model 
on systems up to width six.\cite{White:2003a, Hager:2005, Fehske:2006}

Another class of systems well suited for 2D DMRG are frustrated bosonic models, although there have been few 
DMRG studies of these systems until very recently. 
Bosonic models can be simple to work with numerically, yet are expected to exhibit new kinds of phases and exotic orders not found in 
one dimension. They also have great potential for controlled experimental realizations with cold atoms.
One topic in bosonic systems that has attracted great interest is the possibility of a supersolid, a phase with simultaneous charge
density wave and superfluid order. A supersolid phase was shown to exist in the unfrustrated triangular lattice model, but the order turned
out to be quite weak. However, Jiang et al.\ have demonstrated using 2D DMRG that adding further neighbor hopping to the 
model significantly stabilizes supersolid order.\cite{Jiang:2009a}

A rather different kind of bosonic phase predicted to exist in 2D is known as a bose metal. 
This phase is not captured by a conventional order parameter, but instead by
a pattern of correlations associated with an entire surface of gapless modes (a `bose surface').\cite{Motrunich:2007}
Using DMRG, it has proved possible to identify remnants of the full 2D surface in quasi-1D models with ring-exchange interactions 
on two-leg \cite{Sheng:2008} and four-leg \cite{Block:2010} ladders. 
Motivated by the original construction of the bose metal in terms of fermionic partons,
DMRG calculations have also been used to show that a frustrated 2D fermionic model could enter a Cooper-pair bose metal phase.\cite{Feiguin:2011}
Recently, a rather different type of ordering was observed for a fermionic system on the kagome lattice, where 2D DMRG was used to
study a metal-insulator transition.\cite{Nishimoto:2010}

One final study that does not fit neatly into the above categories but certainly bears mentioning is the investigation by
Jeckelmann and White of the Holstein model, which describes a single fermion interacting with bosonic lattice 
vibrations or phonons.\cite{Jeckelmann:1998}
An effective way of dealing with the large number of boson species in this model was to split 
each site into many smaller ones having fewer degrees of freedom. Using this technique it proved possible to simulate 
cylindrical systems with widths up to twenty!

\section{Techniques for Two Dimensional DMRG \label{sec:techniques}}

When working at the frontier of current numerical capabilities, it is important
to ensure that individual simulations give trustworthy results and then combine
these results correctly to build an accurate picture of a 2D model.  For two
dimensional DMRG, the main obstacle to overcome is that the number of states
kept must be increased exponentially with the width of the system to maintain a
constant accuracy.\cite{Liang:1994}  In practice, this puts an upper bound on
the system sizes that can be simulated. In addition, one should use open or
cylindrical boundary conditions, as opposed to fully periodic,
in order to avoid squaring the number of states required for a given accuracy (see below).

Although the first restriction is unfortunate, the use of cylindrical boundary
conditions is hardly the drawback it is often portrayed to be. In fact,
cylindrical boundaries provide a significant degree of control over simulations
and can be very useful for inferring properties of the full 2D system.  The
limitations of two dimensional DMRG are further mitigated by the enormous
flexibility of the DMRG approach.  DMRG gives full access to the many-body
wavefunction. This means, for instance, that the ground state computed for one
Hamiltonian can be used as input for simulations with a different Hamiltonian,
inviting a range of techniques such as biasing initial states to detect
symmetry breaking or changing a Hamiltonian in mid-simulation to locate phase
boundaries.

In what follows we discuss these techniques and others that have been
successfully used in cutting-edge calculations.  The hope is to equip
practitioners to pursue new state-of-the-art 2D DMRG calculations and to
motivate them to create their own techniques, extending the reach of current
numerical methods.

\subsection{Ensuring Ground State Convergence}

A successful DMRG study of a two dimensional system involves multiple independent calculations for various system sizes over a
range of parameters. In order to deduce the correct behavior of the 2D system from these results, one must guarantee that each ground state 
calculation is well understood and as accurate as possible.
DMRG can fail to find the true ground state for two basic reasons. The first is that the number of states kept after each truncation of the wavefunction may be too small to represent the wavefunction accurately. The second issue is more subtle and arises because while DMRG uses exact diagonalization locally, it 
is globally a variational method and can therefore get stuck in a metastable state.\cite{Dukelsky:1998}

\begin{figure}[tp]
\includegraphics[width=\columnwidth]{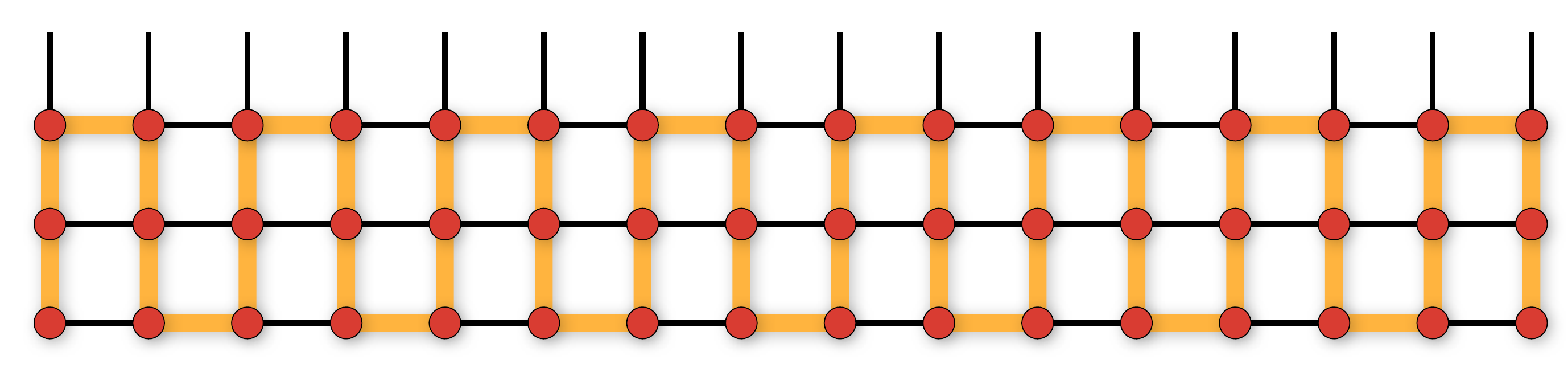}
\caption{Interaction bonds for a nearest-neighbor model with cylindrical boundary conditions on a three leg ladder. 
The thicker bonds indicate interactions which remain nearest-neighbor along the 1D path used by DMRG while thinner bonds are treated as further
neighbor interactions.}
\label{fig:cylindrical}
\end{figure}

There are a number of ways to ensure that one does not over-truncate the
wavefunction and that a sufficient number of states $m$ are being kept.  The
most important is to avoid using fully periodic boundary conditions, as they
require $m^2$ states to represent the same wavefunction requiring only $m$
states with open or cylindrical boundary conditions.\cite{Schollwoeck:2011}
By cylindrical boundary conditions we mean open boundary conditions along the
larger lattice direction (the `length', or $x$ direction) and periodic boundary
conditions along the smaller direction (the `width'). A typical setup using
cylindrical boundary conditions is shown in Figure~\ref{fig:cylindrical}.

After a calculation is under way, the number of states kept should be increased
systematically until the energy and any other observables of interest converge
to a specified tolerance. This can be done by directly adjusting $m$, or by
fixing the maximum truncation error at a step and have the code determine $m$
on the fly. An optimal approach may involve taking both a maximum (and minimum) $m$
into account as well as a requested truncation error.  A minimum $m$ is important
because the reported truncation error early on in the calculation could be very
inaccurate, resulting in slow convergence because of too small an $m$.

Metastability issues are more difficult to deal with systematically.
Practitioners must use a number of checks in addition to physical insight to
verify that the wavefunction has reached the true ground state. For many
models, specific properties of the ground state are known in advance. 
Additionally, if one uses a DMRG code that represents wavefunctions as a matrix product state, then by representing the
Hamiltonian as a matrix product operator it is possible to efficiently compute
the variance $\sigma_H = \bra{\psi} H^2 \ket{\psi} - E_\psi^2$ and confirm that
$\ket{\psi}$ is an eigenstate.

There is a general technique which helps DMRG avoid being stuck in metastable
states. This technique is particularly helpful in making sure the MPS wavefunction includes
correlations coming from terms in the Hamiltonian connecting distant sites in the MPS
path.  For example, a nearest neighbor hopping in a 2D strip may connect sites which are the transverse width apart in the MPS path.  These correlations may have trouble getting started, since the extra states needed to allow the hopping at one of the two sites may not help lower the energy unless the extra states at the other site are already present. A special ``noise'' term can be added to the density matrix at each step which takes into account all terms connecting the left and right blocks.\cite{White:2005} This extra noise helps such long range correlations get started; generally the noise is turned off in later sweeps.  This technique is particularly important if, for some reason, fully periodic boundary conditions must be used.

A key way to avoid metastability is to begin with a wavefunction that is
already close to the true ground state. Good initial wavefunctions may be hard
to come by for an unfamiliar model; here careful calculations on smaller system
sizes or at smaller values of $m$ where DMRG has better control can be an
excellent guide. Such calculations allow one to identify the dominant
correlations within the ground state which may otherwise be obscured by strong
fluctuations on larger lattices or for larger $m$. 

For a system which is expected to have a conventional symmetry breaking ground
state, such as an antiferromagnet on a bipartite lattice, a N\'eel state may be
a sufficiently good starting point.  For systems with more subtle order, the
initial wavefunction can be produced by starting with a lower symmetry or
`pinned' Hamiltonian. For example, if a system is expected to have a ground
state with valence bond solid order, the Hamiltonian can be modified by adding
pinning fields $\lambda\, \vec{S}_i \cdot \vec{S}_j$ for each pair of sites
$i,j$ connected by a valence bond.  Then, after a few sweeps $\lambda$ can be
gradually tuned to zero allowing the system to relax to its true ground state.

Using an initial state or a pinning field can also be helpful for ruling out
hypothesized properties of a model.  If DMRG restores a symmetry explicitly
broken by the initial state, one has strong evidence against that particular
ordering scenario.  This method has been used to rule out a type of
checkerboard order for the $t-J$ model \cite{White:2004} and more recently as
evidence against a particular type of valence bond solid order for the kagome
Heisenberg antiferromagnet.\cite{Yan:2011}

When dealing with a complex lattice or a phase with a large unit cell, one
way to deal with metastability issues and minimize the number of
states needed is to experiment with multiple DMRG paths. For a fixed value of
$m$, DMRG is better able to capture entanglement on Hamiltonian bonds that
remain nearest-neighbor when mapped to 1D. Choosing the DMRG path judiciously
can even permit complex initial wavefunctions such as valence bond solids to be
represented exactly with only a small value of $m$.  Having the ability to
reproduce the same ground state with different DMRG paths can also provide strong
evidence that one has found the true ground state and not a metastable
solution. 

After having gained a good understanding of smaller systems, one wants to push
DMRG calculations up to the largest accessible widths. At these widths, there
is less control, so in order to produce accurate results it is
 very useful to extrapolate from more controlled limits.  For DMRG, a natural
extrapolation parameter is the truncation error $\varepsilon$ (the sum of discarded density
matrix eigenvalues). The energy has long been extrapolated to zero truncation error,
where normally a linear extrapolation of $E$ versus $\varepsilon$ is best.
Remarkably, within a DMRG calculation local measurements performed on the two central
sites at each step also have errors varying linearly with $\varepsilon$!  This
is one reason why measuring local quantities, perhaps in response to a perturbation,
is usually preferred to correlation functions, whose error varies as $\varepsilon^{1/2}$.\cite{White:2007}
Efficient extrapolations can be performed 
using results from a single DMRG calculation with increasing $m$, but it is important to
repeat each $m$ for two full sweeps (and to extrapolate using the last of the four half-sweeps)
to ensure that the calculated $\varepsilon$ is consistent
enough for extrapolation.

\begin{figure}[tp]
\includegraphics[width=\columnwidth]{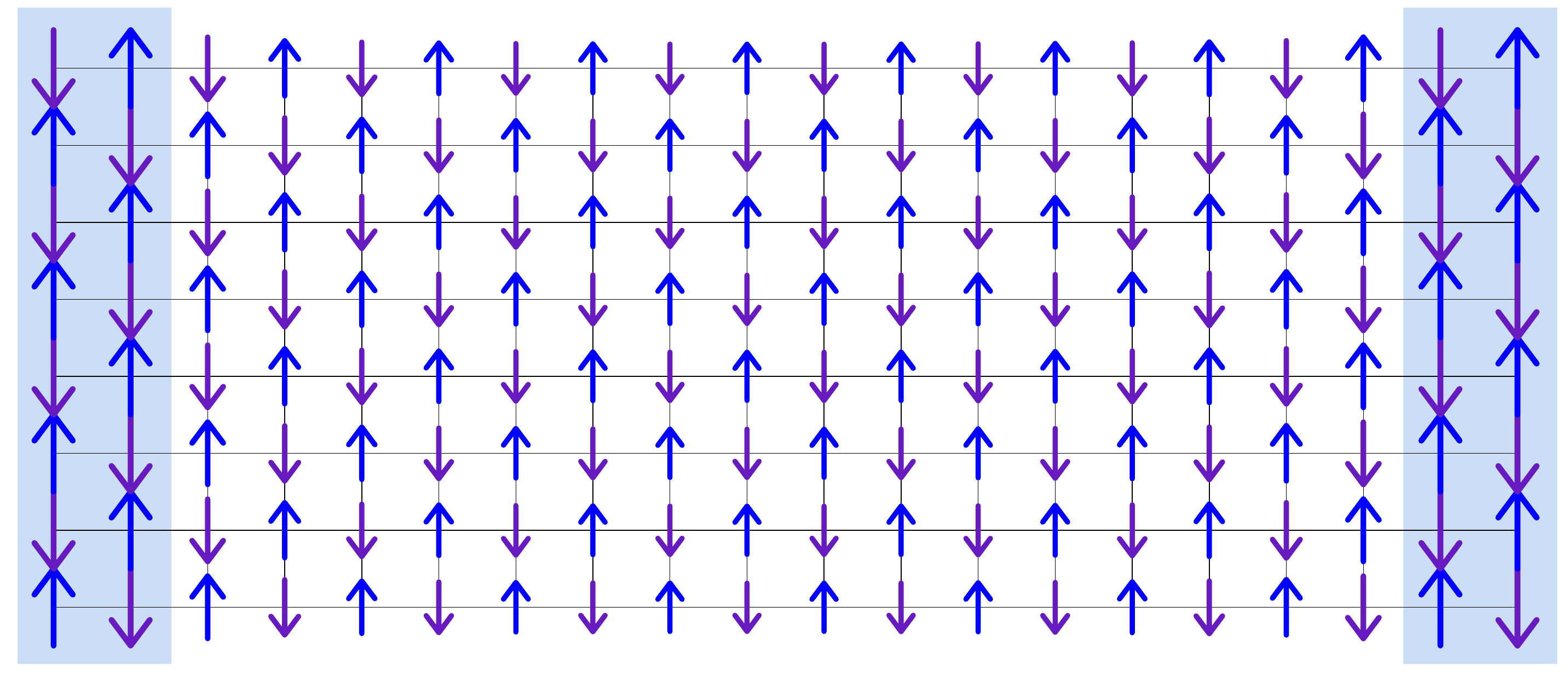}
\caption{Results of a DMRG calculation for the Heisenberg model on a 16$\times$8 cylinder with antiferromagnetic order pinned at the open boundaries.
To work in the strong pinning limit, it is useful to imagine the finite system embedded within a larger system 
acted on by an infinitely strong field (shown here as the shaded regions). The pinning fields at the physical edges are  
determined by the Hamiltonian bonds connecting the 
real and fictitious system.}
\label{fig:boundary}
\end{figure}

The flexibility of DMRG even allows the use of other extrapolation parameters that may work
better than the truncation error in certain cases.  For example, one can add a small
perturbation $\lambda\,H^\prime$ to the Hamiltonian and extrapolate the energy
in $\lambda$.  For this approach to work well, the ground states of the
perturbed Hamiltonian should be less entangled than the true ground state.
Furthermore, by choosing $H^\prime$ to have a vanishing expectation value with
respect to the ground state of $H$, the first derivative of the energy with
$\lambda$ can be tuned to zero, increasing the accuracy of the extrapolation.\cite{Yan:2011}

\subsection{Working Around Finite Size Limitations \label{sec:fs} }

A variety of approaches
can be taken to predict bulk 2D behavior from sets of finite systems - here we discuss
some that are particularly useful in the context of DMRG. 
Most of these approaches utilize cylindrical boundaries.  On the two open edges of the cylinder, one is
free to apply local fields (``pinning" it), or to perturb it in other ways,
in order to make the bulk represent 2D most accurately.
A favorable side-effect of applying a boundary pinning field may be a reduction of entanglement,
improving the DMRG convergence.  For example, an antiferromagnet on a finite system
typically has a singlet ground state, but one may regard it as a superposition of 
antiferromagnetically ordered states with different directions for the order parameter.
Pinning can select one order parameter direction, reducing the complexity and entanglement
of the state, while simultaneously representing the broken-symmetry 2D properties more
faithfully.\cite{White:2007}

It can be very useful to choose strongly pinned edges.  To choose this type of pinning,
it is helpful to think of the finite cylindrical system as part of an infinite cylinder.  A fictitious
infinitely strong pinning field is applied to the fictitious system outside the finite part.
The fictitious part then acts on the boundary of the real system through the Hamiltonian terms
connecting the two. Figure~\ref{fig:boundary} shows the results of a ground
state calculation for the $S=1/2$ Heisenberg model on a cylinder of width 8 and
length 16. Two rows of the fictitious system are also shown at each end of
the cylinder. The magnetization is enhanced near the pinned sites, then
approaches the 2D bulk value at the center (if the aspect ratio is chosen
properly as discussed below).

After gaining a good understanding of the system's ground state properties, the next step is to simulate 
multiple system sizes to build up an accurate picture of the thermodynamic limit. 
Because DMRG scales much more favorably with the length of a system than with its width,
it is a good idea to group results at fixed width together. Naively then, it would seem that the best 
approach is to simulate the longest possible system at each width. 
But there are much more efficient ways to proceed. 

\begin{figure}[bp]
\includegraphics[width=\columnwidth]{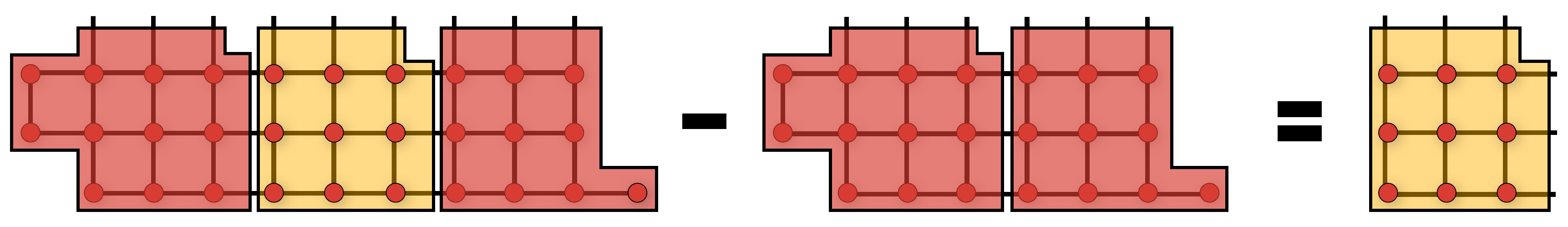}
\caption{Properties at open edges are not typical of the bulk model and may include pinning fields or extra sites designed to accommodate a 
certain ground state. 
Subtracting energies calculated for two finite length system cancels the leading edge effects giving values that rapidly converge to the properties
of the infinite cylinder. }
\label{fig:subtraction}
\end{figure}

For a fixed width, the energy per site of a length $L_x$ system approaches the infinite
system value with an error proportional to $1/L_x$ because of a constant term in the energy
from the open ends. This slow convergence reduces
the efficiency and accuracy of direct extrapolations.  Therefore, it is convenient to
determine bulk cylinder energies by subtracting the energies of different finite-length cylinders.
As illustrated in Figure~\ref{fig:subtraction}, subtracting cancels edge
effects, leaving only the bulk energy of the larger system, which rapidly
converges to the infinite value as a function of $L_x$.  The convergence rate
depends on the bulk correlation length(s).  If the correlation lengths
are infinite, one can extrapolate the subtracted energies with a polynomial for
the highest accuracy. 

\begin{figure}[tp]
\includegraphics[width=\columnwidth]{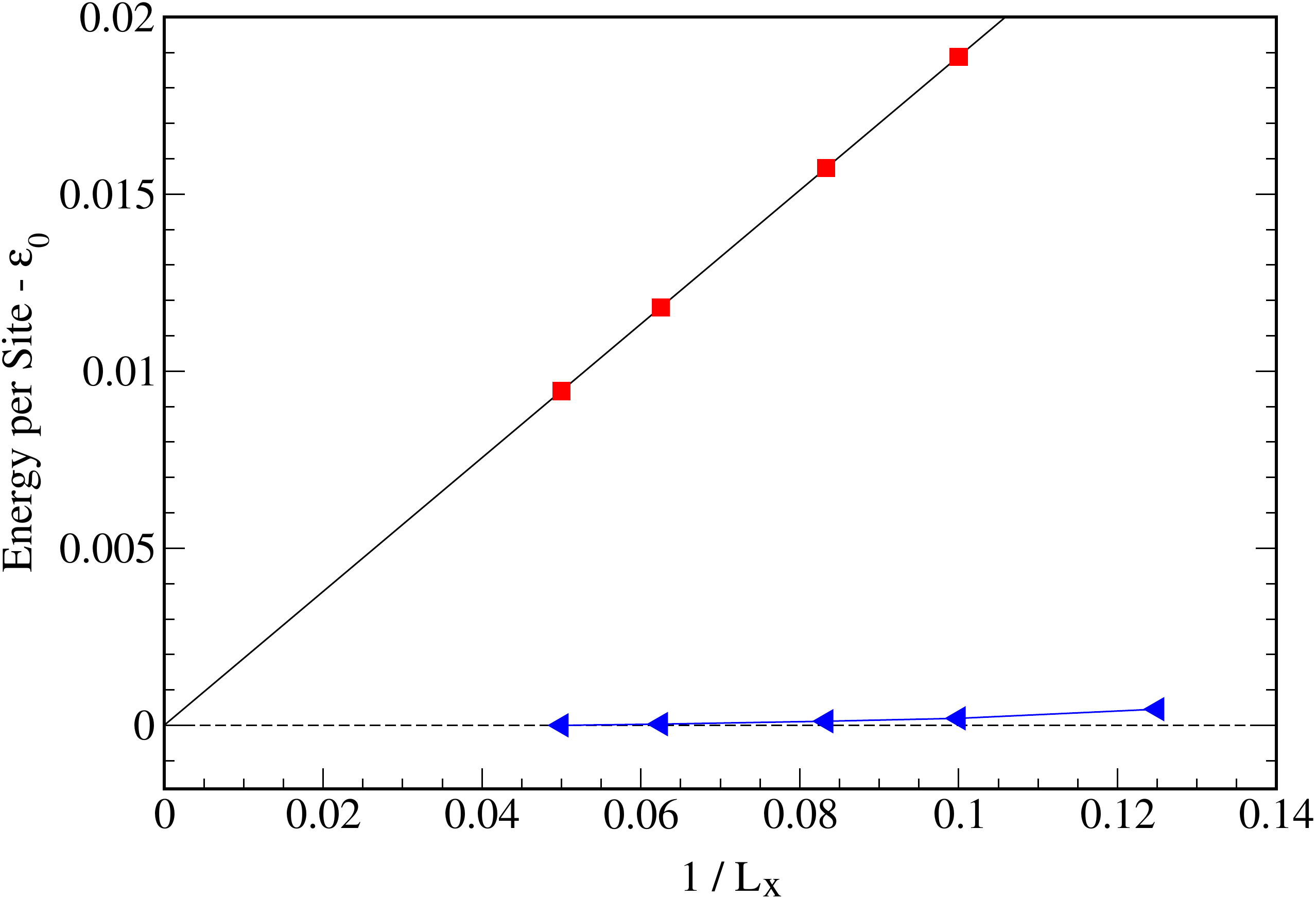}
\caption{Subtraction method for estimating the energy per site of the $S=1/2$ Heisenberg model on an infinite cylinder of width $L_y=6$.
The upper points (squares) are the energy per site of finite systems with $L_x=8,10,12,16\ \text{and}\ 20$ - error bars were smaller than the symbol size. 
The lower points (triangles) are found by subtracting the energy of the length $L_x/2$ system from the length $L_x$ system and dividing by the number of extra sites 
of the larger system. The energy estimate $\epsilon_0 = -0.67172(7)$ for the infinite cylinder is taken from the best subtracted data point.}
\label{fig:subtracted_energies}
\end{figure}

As an example of the subtraction method, Figure~\ref{fig:subtracted_energies} show results for the
square lattice $S=1/2$ Heisenberg model on 6 leg ladders of length $L_x=8$
through $L_x=20$. The  bare energies of the finite ladders have a strong
$1/L_x$ dependence, but the subtracted energies rapidly converge to a constant
value and require no fine tuning of the boundary conditions.  While one has to
extrapolate the single-cylinder data, the subtracted energies converge quickly
enough that the best point can be used as the infinite cylinder energy
estimate.

\begin{figure}[t]
\includegraphics[width=\columnwidth]{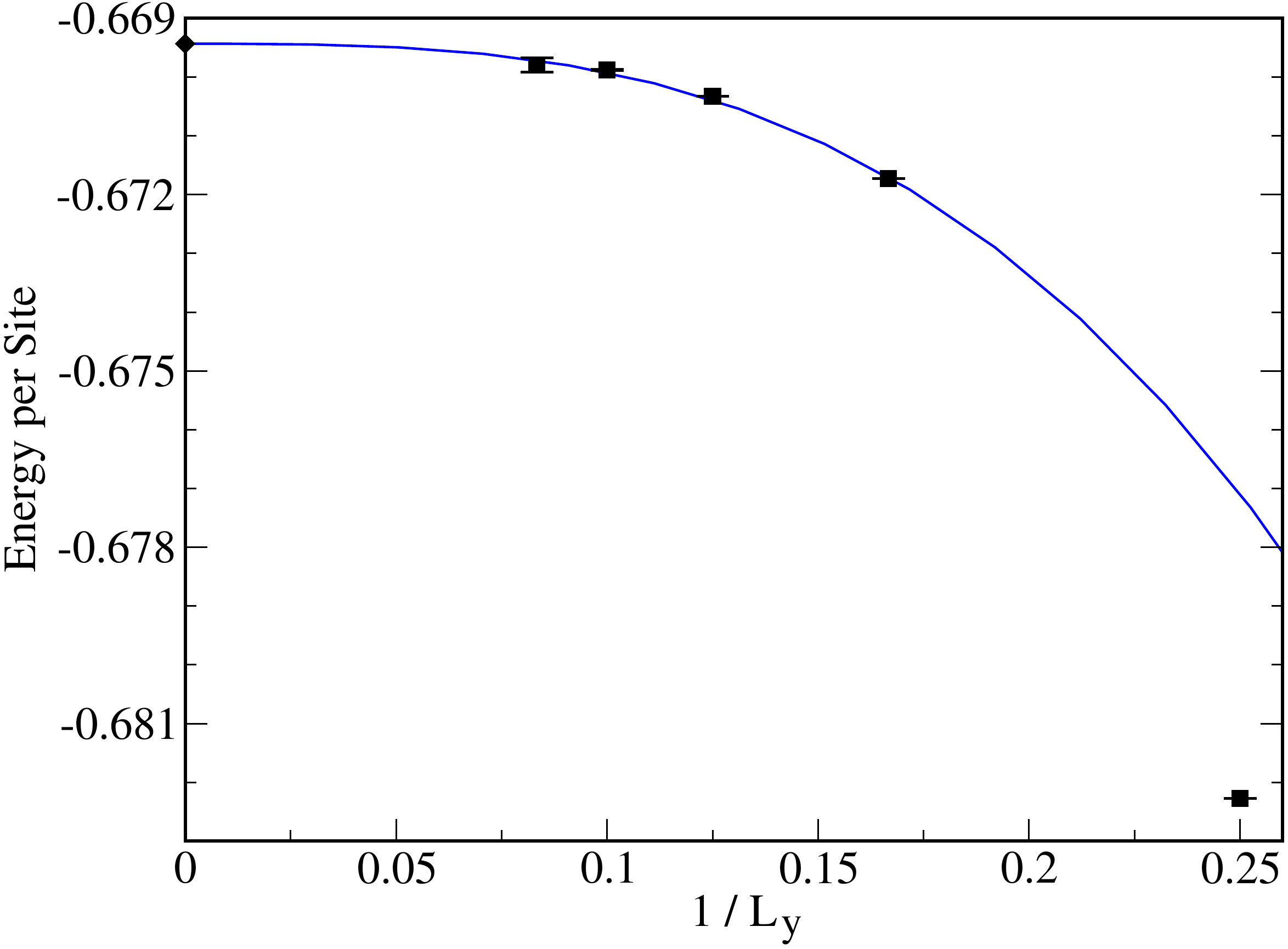}
\caption{Extrapolation of energy estimates for infinite cylinders of the $S=1/2$ Heisenberg model with $J=1$.
The energy densities for widths $L_y=6,8,10$ and $12$ are fit to a cubic polynomial $a+b\:L_y^{-3}$ (see Ref.~\onlinecite{Neuberger:1989}),
giving an energy estimate $\epsilon_0=-0.6694(3)$ for the 2D system.}
\label{fig:extrap}
\end{figure}

Finally, after obtaining the bulk energies of infinite cylinders with various widths $L_y$, one
can extrapolate in $L_y$ to estimate the 2D energy. For the case of periodic boundary conditions---applicable for the infinite cylinders---
the leading finite-size corrections to the energy density of the Heisenberg model are expected
to vary as $1/L_y^{3}$.\cite{Neuberger:1989} In Figure~\ref{fig:extrap} we show the result of fitting 
the infinite cylinder energies with $L_y\geq 6$ to this form. We obtain an estimate  $\epsilon_0=-0.6694(3)$
consistent with the best published Monte Carlo result $\epsilon_0=-0.669437(5)$.\cite{Sandvik:1997}

Bulk order parameters should be obtained differently from energies---considering
infinite cylinders is not optimal.  Asymptotically, infinite length cylinders 
of constant width will exhibit 1D behavior.

\begin{figure}[t]
\includegraphics[width=\columnwidth]{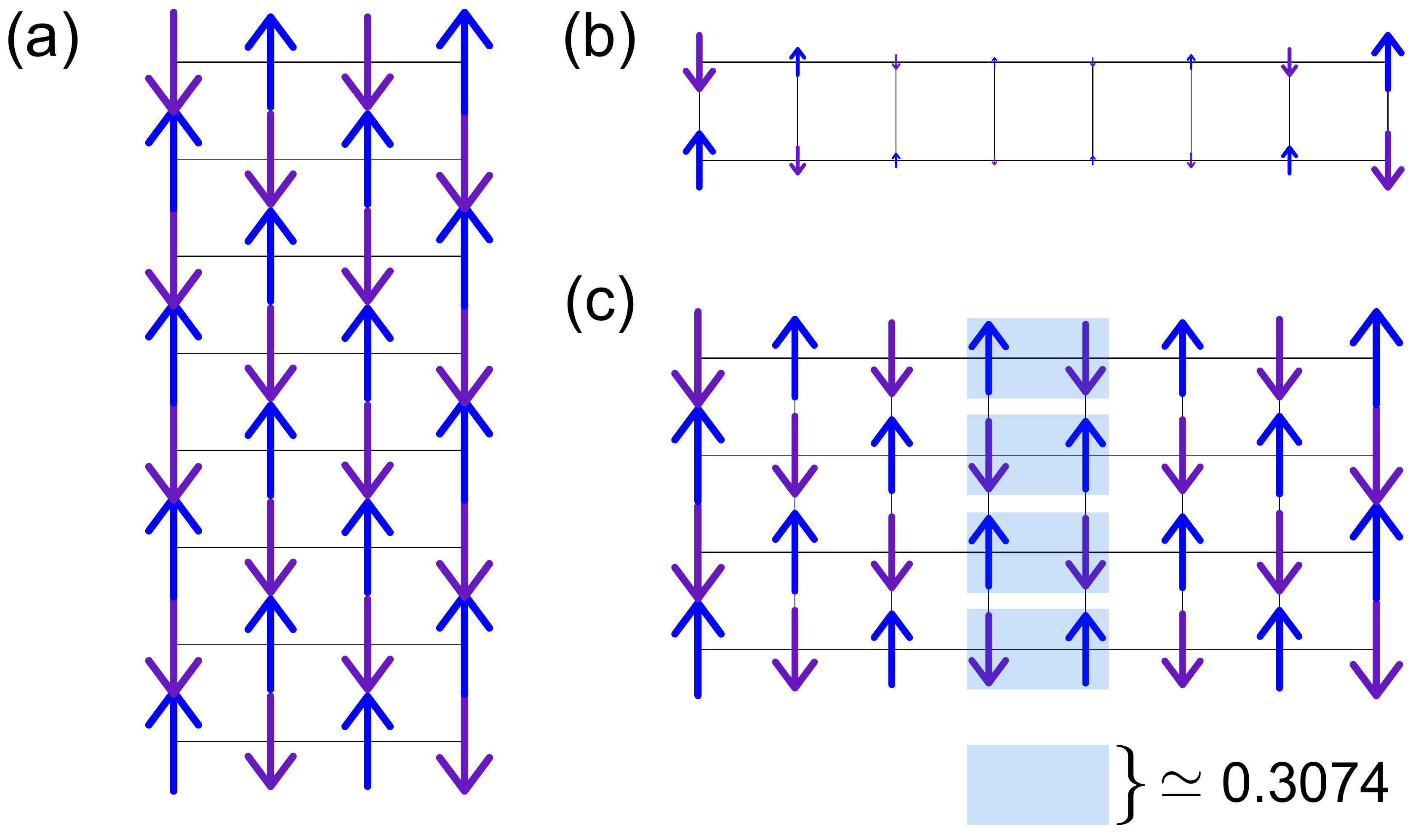}
\caption{Ground states of the $S=1/2$ Heisenberg model on  (a) 4$\times$8, (b) 8$\times$2 and (c) 8$\times$4 cylinders. 
All cylinders have a staggered pinning 
field of strength $J/2$ applied at the open ends. Because cylinder (a) is too short, edge effects push its bulk magnetization
above the true 2D value. Cylinder (b) is too one-dimensional so its bulk magnetization falls below the 2D value. Cylinder (c) has an
aspect ratio of $2\!\!:\!\!1$ which is almost ideal,\cite{White:2007} making its bulk a good approximation of the infinite 2D system.
The magnetization of cylinder (c) at the center rungs ($m\simeq0.28$) is close to the estimated 2D value of $m\simeq0.30743$.\cite{Sandvik:2010L}}
\label{fig:aspect}
\end{figure}

A particularly effective approach for measuring order parameters is to try to choose an optimal aspect ratio
$\alpha=L_x/L_y$ for finite cylinders. The approach utilizes the strongly pinned boundary conditions
described above.  Say we decide
to extrapolate to 2D only using cylinders with a fixed $\alpha$. 
As illustrated in Figure~\ref{fig:aspect}, if $\alpha$ is too
small, edge effects suppress fluctuations and our estimates will approach the
2D value from above; if $\alpha$ is too large the physics resembles that of a
1D chain and our estimates approach from below. In several ``typical" systems and in
associated continuum theories it
has been found that there is an
optimal ratio $\alpha^*$ where leading corrections in $1/L_y$ vanish.\cite{White:2007}
Extrapolations performed at fixed $\alpha^*$ will be nearly flat, improving
their accuracy.  For periodic boundary conditions the ideal aspect ratio turns
out to be about $\alpha^* \simeq 7$. For cylindrical boundaries it is much
lower---around $\alpha^* \simeq 1.9$.\cite{White:2007} This is yet another
reason why cylindrical boundary conditions should be preferred.  Finite
size effects can be drastically reduced even if one does not use the exact
aspect ratio.

\subsection{Gaps and Excited States}

In addition to ground state properties, one usually wants to have a good understanding of the elementary excitations of a system. 
Calculating the energy gap to the first excited state is of fundamental importance for classifying phases and estimating their
robustness to perturbations. The excited states themselves can be useful for understanding subtle orders present in the ground 
state and for computing quantities that are accessible to experiments.

Traditionally there have been two ways to find excited states and gaps using DMRG and both methods 
are a good option for studying 2D systems.
The simplest situation is when an excited state lies in a different quantum number sector than the ground state.
This is the case when calculating the spin gap of a magnet, for example.
By taking advantage of quantum numbers DMRG can directly target the ground state of an excited sector.
This method is preferred as long as the 
excitation is not attracted to the open ends of the system, which can easily be checked from measurements
of local quantities. If the excitation is attracted to the ends, the restricted sweeping approach described below
is a good option.

When studying excited states in the same sector as the ground state, such as singlet excitations, or 
when studying a model with no conserved quantities one can simultaneously target the lowest few eigenstates. 
In this scheme the superblock wavefunctions for each state are kept separate but they share one 
set of boundary blocks. It is often necessary therefore to keep an increased number of states in order to approximate 
different wavefunctions using a single truncated basis.

But there are situations where neither of the above approaches is a good option. 
Models of topological phases often have no conserved quantities plus large ground state
degeneracy that makes multiple targeting inefficient. 
For cases like these, it is helpful to take advantage of the flexibility offered through 
using matrix product states (MPS) and matrix product operators.
Representing a state as an MPS gives DMRG access to the entire wavefunction and makes it possible
to compute overlaps between wavefunctions found through separate calculations.

This flexibility gives  a new way to find excited states as follows. First, DMRG is used to compute 
a ground state $\ket{\psi_0}$ of the Hamiltonian $H$ as an MPS. Then, one defines a Hamiltonian
$H^\prime = H + w P_0$ where $P_0=\ket{\psi_0}\bra{\psi_0}$ is a projection operator and $w$ is an energy penalty for states
not orthogonal to $\ket{\psi_0}$. If $w$ is large enough, the ground state $\ket{\psi_1}$ of $H^\prime$ will be the second lowest eigenstate of $H$
(its first excited state or a second ground state).

\begin{figure}[b]
\includegraphics[width=\columnwidth]{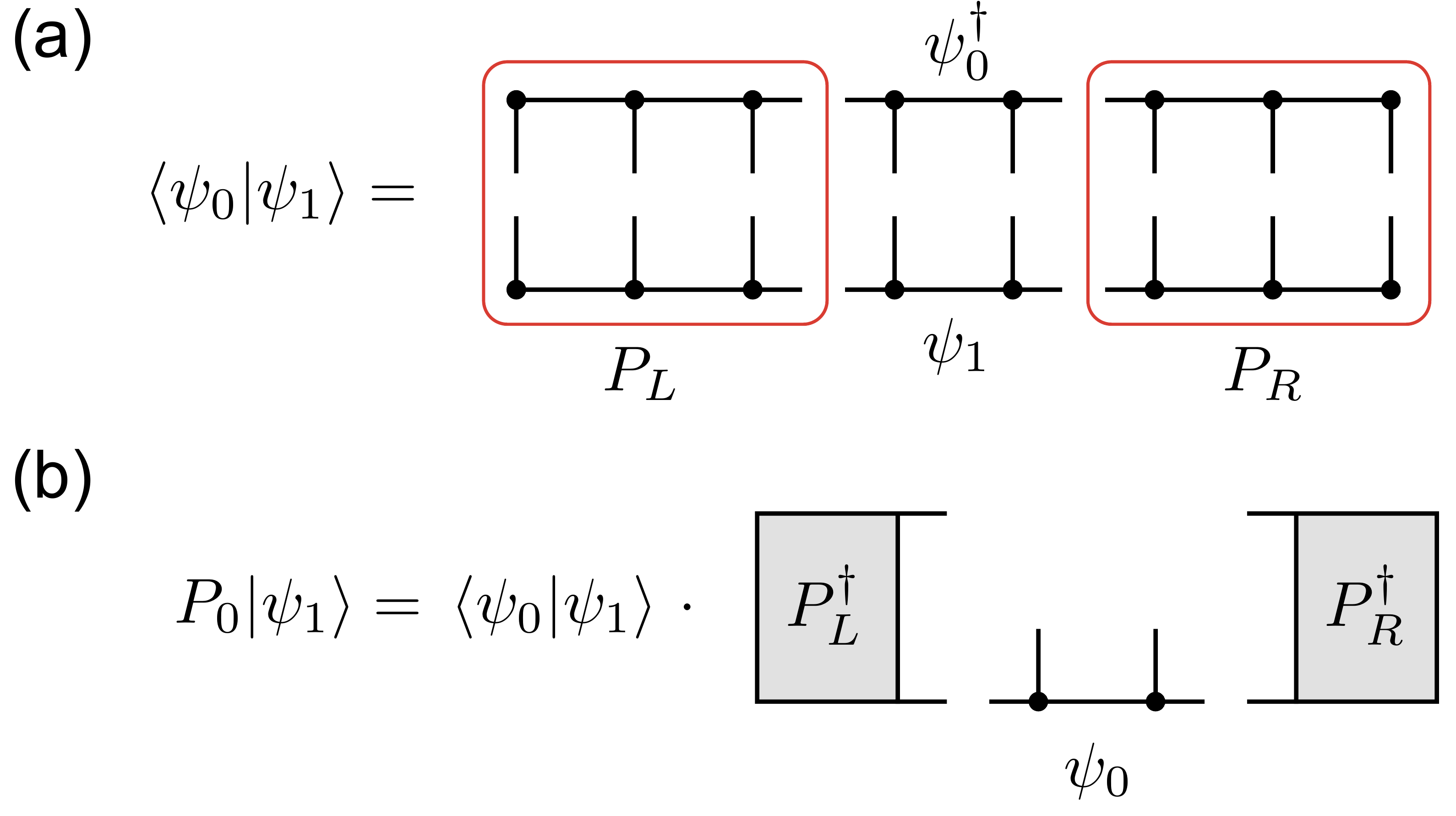}
\caption{Steps for computing the product of a projector $P_0=\ket{\psi_0}\bra{\psi_0}$ with a state $\ket{\psi_1}$ in DMRG. After computing the 
tensors $P_L$ and $P_R$ that project $\bra{\psi_0}$ into the current basis, one (a) computes the overlap $\bracket{\psi_0}{\psi_1}$ then 
(b) multiplies it by the projected form of $\ket{\psi_0}$. }
\label{fig:ortho_dmrg}
\end{figure}

\begin{table*}[htp]
\caption{Comparison of Some Leading Numerical Methods for 2D Strongly Correlated Systems}
\begin{center}
\begin{tabular}{ p{2.55cm} p{3.2cm} c c p{2.7cm} c  p{3.cm}}
\hline \hline
\bf Method & \bf Approach & \bf Variational & \bf 2D? & \bf Sign Problem & \bf Biased & \bf Typical Computational Effort \\ \hline
\mbox{Worldline QMC} \cite{Sandvik:2010} & Statistical Sampling & No &Yes & Yes & No & N \\ \\
Determinantal QMC \cite{Blankenbecler:1981,White:1989} & Statistical Sampling & No &Yes & Yes & No & N$^3$ \\ \\
Variational QMC \cite{Foulkes:2001} & Statistical Sampling & Yes & Yes & \mbox{Fixed by Guessed} Wavefunction & Yes & N$^3$ \\ \\
\mbox{Series Expansion} \cite{Domb:1974} & Extrapolated \mbox{Taylor Series} & No & Yes & No & Yes & $<$ 10 or 20 Terms \\ \\
DMRG & Low Entanglement & Yes & Width $\approxlt 12$\ \ & No & Very Slight$^{a}$ & $m^3$ \\ \\
PEPS & Low Entanglement & No$^b$ & Yes & No & Very Slight$^a$ & $D^{10}$--$D^{12}$ (Ref.~\onlinecite{Corboz:2011pc}) \\ \\
MERA & Low Entanglement & Yes & Yes & No & Very Slight$^{a,c}$ & $D^{16}$ (Ref.~\onlinecite{Corboz:2010f}) \\ \\
\hline
\end{tabular}
\end{center}
\begin{flushleft}
$a$. Indicates a bias toward states having low entanglement. \\
$b$. Though PEPS themselves are variational, observables must be computed through a controlled approximation. \\
$c$. The pattern of tensors chosen may favor a certain type of ground state.
\end{flushleft}
\label{table:methods}
\end{table*}

But when finding $\ket{\psi_1}$ it is neither necessary nor advisable to compute $H^\prime$.
One should instead work with $H$ and $P_0$ separately. During a normal Lanczos or Davidson step in DMRG, the product $H \ket{\psi_1}$ gets computed
in the current local basis. The product $P_0 \ket{\psi_1}$ can be computed in a similar fashion following the procedure in Figure~\ref{fig:ortho_dmrg}. 
The resulting tensors $H\ket{\psi_1}$ and $w P_0 \ket{\psi_1}$ can then be added to form $H^\prime \ket{\psi_1}$.


Having found $\ket{\psi_1}$, one can go on to compute the next excited state by including both $P_0$ and $P_1=\ket{\psi_1}\bra{\psi_1}$ 
in the effective Hamiltonian. 
Many low-lying states can be found this way with a cost that is quadratic in
the number of states, although the method can only be pushed so far 
unless the previous wavefunctions are determined to high accuracy.
Finally, after finding a set of low-lying states $\ket{\psi_n}$
it is possible to obtain an even more accurate spectrum by computing the eigenvalues of $H^{m,n} = \bra{\psi_m}H\ket{\psi_n}$.
This last step rotates away any remaining non-orthogonality present in the states $\ket{\psi_n}$ ensuring that energies are computed in 
an optimal basis.

Each of the three methods discussed above has its merits, but one issue that affects them all is the possibility of spurious excitations
at the open ends of the cylinder. In most cases such an excitation is uninteresting since it has no analogue 
in the infinite 2D system. If this problem arises, one way to deal with it is through restricted sweeping. First, the ground state
is computed as usual to very high accuracy. Then, within the same calculation one switches over to the first excited state
either by changing quantum numbers or through multiple targeting. But now the sweeping range is restricted only to the bulk of the cylinder, 
keeping the system fixed to the ground state basis near the open ends.  By mixing with this fixed basis, excitations can have tails 
extending beyond the sweeping region, but will not get stuck in an unphysical state where they live only at an edge.

\section{Two Dimensional Tensor Network Methods \label{sec:tns}}

Based on DMRG's success for 1D and quasi-1D systems, it is reasonable to ask if DMRG can be extended to 2D in a more natural way.
Early attempts to do so failed because they did not take into account the essential differences in entanglement scaling of 1D and 2D systems.
Ground states of 1D gapped Hamiltonians obey an area law: any subsystem is entangled with its environment only through the boundary
connecting the two. This means that ground states of gapped 1D systems can be represented accurately by matrix product states
with a fixed bond dimension.\cite{Hastings:2007}

Most---if not all---gapped phases in 2D also obey an area law, but in 2D this means that the entanglement entropy of a subregion grows proportionally to its 
linear size. To capture this effect with an MPS requires that its bond dimension  increase exponentially with the system size no matter how it is 
embedded into a 2D lattice. The key, then, to obtaining a scalable version of 2D DMRG is to identify new classes of variational wavefunctions
that extend matrix product states to reproduce 2D area law behavior.

Though there have been a number of promising attempts in this direction, two classes of wavefunctions stand out based on their usefulness in
simulating a number of realistic 2D models. The first is a natural extension of an MPS known as a Projected Entangled Pair State or 
PEPS.\cite{Nishino:2001,Verstraete:2004} Just as an MPS can be viewed as a network of rank 3 tensors, a PEPS is a network of rank 
$Z+1$ tensors with $Z$ the coordination number of the lattice. 
The usefulness of PEPS is somewhat limited by the steep computational cost required to optimize them, which grows as $D^{10}$ where $D$ is the bond dimension.
And although a PEPS is a well-defined variational wavefunction, it is necessary to introduce approximations when computing observables, leading to
energy estimates that are no longer variational.
But because of the higher coordination number of 2D lattices, one expects that 2D ground states are already represented quite well by a PEPS with 
\mbox{$D\sim10$--$100$} versus an MPS for a 1D system where $m$ must be many hundreds to achieve the same accuracy.\cite{Verstraete:2008}
Perhaps the most attractive feature of the PEPS approach is its ability to work directly in the thermodynamic limit using the 
so-called iPEPS method.\cite{Jordan:2008,Orus:2009c} The iPEPS method has been successfully applied to a wide variety of models including 
orbital models, \cite{Orus:2009} frustrated magnets, \cite{Bauer:2009} interacting bosons \cite{Jordan:2009} and interacting fermions.\cite{Corboz:2010b,Corboz:2010}

\begin{figure}[t]
\includegraphics[width=\columnwidth]{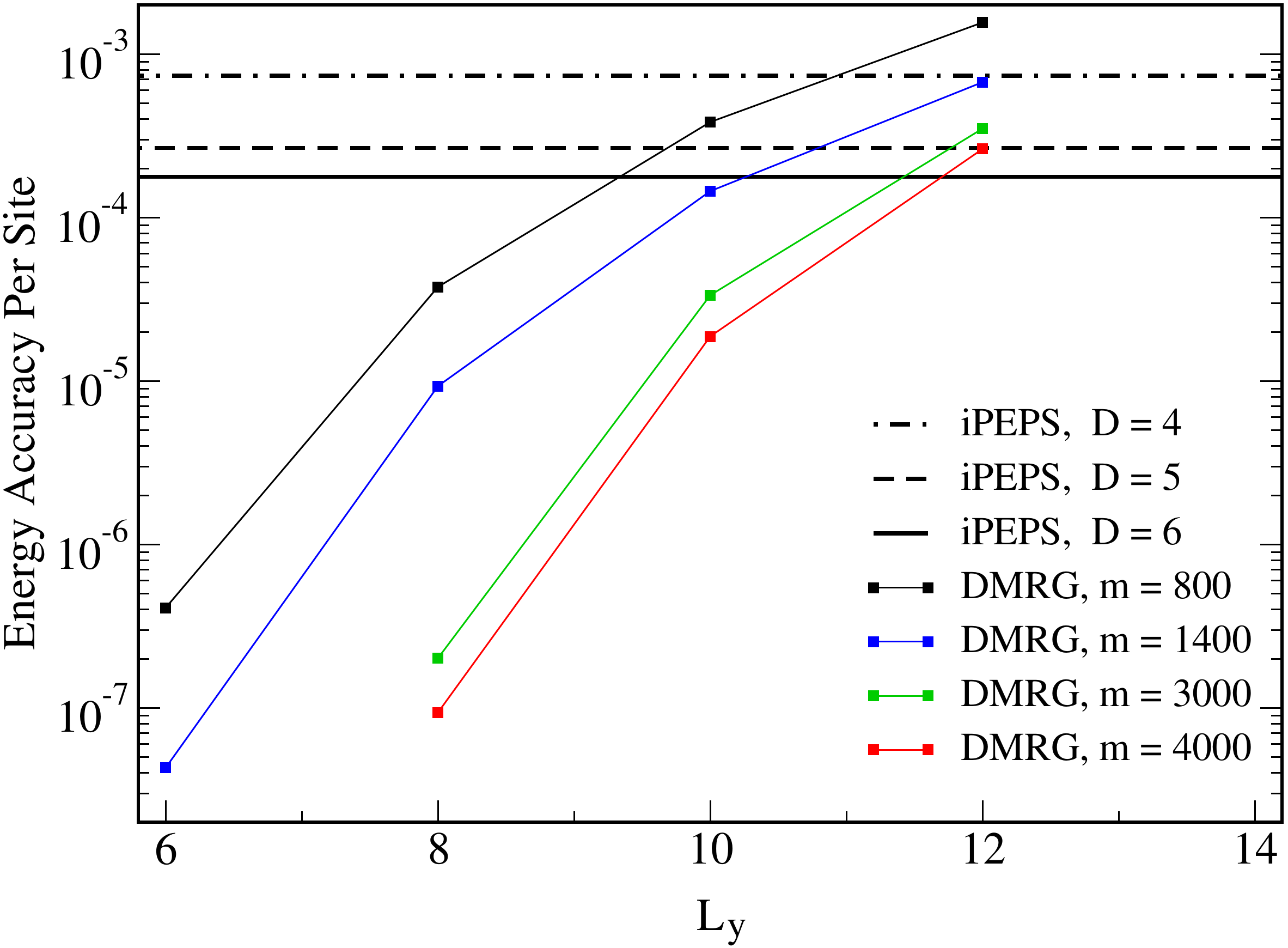}
\caption{Plots of the accuracy achievable with DMRG for cylinders of width $L_y$ and a maximum number of states kept $m$. Data is 
for the $S=1/2$ square lattice Heisenberg model with $J=1$. The DMRG accuracy is defined as the difference between the energy for a fixed $m$ and the energy obtained by extrapolating in the truncation error.
The horizontal lines show the difference between iPEPS energy estimates \cite{Corboz:2011pc} and the QMC estimate $\epsilon_0=-0.669437(5)$.\cite{Sandvik:1997}
Because one is always restricted to some 
maximum $m$ in DMRG, it is only worthwhile to extrapolate using widths $L_y$ for which there is a better accuracy than iPEPS; for simulations of 
larger systems it is more sensible to use PEPS or MERA directly. }
\label{fig:crossover}
\end{figure}

A second type of wavefunction that has proven effective for 2D is the Multiscale Entanglement Renormalization Ansatz (MERA). 
Originally conceived as a concrete realization of Kadanoff's real space renormalization group, a MERA generalizes an MPS
by extending the 1D chain of tensors into a layered structure where each layer represents a coarser
length scale in the RG process.\cite{Vidal:2007,Evenbly:2009} The MERA approach can naturally be extended to 2D 
by using tensors that group patches of lattice sites together instead of 1D segments. 
But unlike the PEPS ansatz, there is no unique MERA for a given lattice. This makes the method more flexible but can
also be a source of bias if the tensors are chosen to favor a certain type of ground state.
MERA optimization techniques currently suffer from a higher computational cost than PEPS (see Table~\ref{table:methods}) 
but a key advantage of MERA over PEPS is that local observables may be computed exactly and efficiently, 
allowing MERA simulations to remain variational.
This property of the MERA comes from its construction in terms of isometries that cancel outside of a `light cone' emanating from the location of 
the operator to be measured.\cite{Evenbly:2009}
Finally, like the PEPS method, the MERA approach can give results for both finite systems and the thermodynamic limit.
Recent applications of MERA to 2D systems include studies of interacting fermions, \cite{Corboz:2010f} orbital systems, \cite{Cincio:2010} 
topological models \cite{Tagliacozzo:2011} and frustrated magnets.\cite{Evenbly:2010}

A key question is this:  when should one use 2D DMRG methods, and when should one try
PEPS or MERA (or another approach)?  As a first step in addressing this issue, we provide in Figure~\ref{fig:crossover} a simple
comparison between iPEPS results \cite{Corboz:2011pc} and DMRG data for the 2D square lattice Heisenberg model. 
(We hope that comparable data for MERA can be provided in the future.)  The accuracy
of iPEPS depends on the tensor dimension $D$, while that of DMRG depends on $m$ and the width
of the cylinder studied.  At some width, which depends on how large an $m$ or $D$ can be treated with
current computing resources, DMRG becomes less accurate than iPEPS.  (Neither the values of $m$
nor of $D$ shown in the figure should be thought of as reflecting the state of the art.)
Whether one should use
DMRG or iPEPS depends primarily on whether the bulk behavior can
be extrapolated with cylinders smaller than this width. 
Alternatively, it can be very useful to use both DMRG and iPEPS or MERA, and compare results.

We would like to acknowledge helpful discussions with Sasha Chernyshev, 
Salvatore Manmana and Philippe Corboz, who was kind enough to provide iPEPS data for the square lattice
Heisenberg model. We acknowledge support from the NSF under DMR-0907500.

\bibliographystyle{2ddmrg}
\bibliography{2DDMRG}

\renewcommand{\tabcolsep}{0.05cm}
\renewcommand{\arraystretch}{1}

\end{document}